# Dark state of electrons in a double two-level quantum system

Yoonah Chung[1,6], Minsu Kim[1,6], Yeryn Kim[1,6], Seyeong Cha[1], Joon Woo Park[1], Jeehong Park[1], Yeonjin Yi[1], Dongjoon Song[2], Jung Hyun Ryu[3], Kimoon Lee[3], Timur K. Kim[4], Cephise Cacho[4], Jonathan Denlinger[5], Chris Jozwiak[5], Eli Rotenberg[5], Aaron Bostwick[5] & Keun Su Kim[1*]

[1]Department of Physics, Yonsei University, Seoul, Korea.
[2]Stewart Blusson Quantum Matter Institute, University of British Columbia, Vancouver, BC V6T 1Z4, Canada.
[3]Department of Physics, Kunsan National University, Gunsan, Korea.
[4]Diamond Light Source, Harwell Campus, Didcot, UK.
[5]Advanced Light Source, E. O. Lawrence Berkeley National Laboratory, Berkeley, CA, USA.
[6]These authors contributed equally: Yoonah Chung, Minsu Kim, Yeryn Kim.
[*]e-mail: keunsukim@yonsei.ac.kr

**A quantum state of matter that is forbidden to interact with photons and is therefore undetectable by spectroscopic means is called a dark state. This basic concept can be applied to condensed matter where it suggests that a whole band of quantum states could be undetectable across a full Brillouin zone. Here we report the discovery of such condensed matter dark states in palladium diselenide as a model system that has two pairs of sublattices in the primitive cell. By using angle-resolved photoemission spectroscopy, we find valence bands that are practically unobservable over the whole Brillouin zone at any photon energy, polarisation, and scattering plane. Our model shows that two pairs of sublattices located at half-translation positions and related by multiple glide-mirror symmetries make their relative quantum phases polarised into only four kinds, three of which become dark due to double destructive interference. This mechanism is generic to other systems with two pairs of sublattices, and we show how the phenomena observed in cuprates, lead-halide perovskites, and density wave systems can be resolved by the mechanism of dark states. Our results suggest that the sublattice degree of freedom, which has been overlooked so far, should be considered in the study of correlated phenomena and optoelectronic characteristics.**

There are many fundamental concepts described by dark across a range of research fields from dark matter in the universe down to dark states in atoms or molecules[1-5]. The dark state refers to the quantum state that does not absorb or emit photons, so that it is undetectable by a spectroscopic means. It is nonetheless unignorable, because this hidden state could be a missing piece of information that is critical to understand quantum phenomena that have remained elusive. It is therefore fundamentally important to identify the existence of other hitherto unknown dark states in nature and to disclose their underlying mechanisms.

The mechanism of atomic, molecular, and excitonic dark states is essentially based on either conservation of angular momentum or quantum interference[1-4]. Even for condensed matter, there are other sources of quantum interference, which are more than one atom of the same kind in the primitive cell, namely, the sublattice. The importance of sublattices in condensed matter has been increasingly recognised, since the experimental discovery of graphene[6], where there are two sublattices typically indexed by A and B. It can be viewed as a two-level quantum system described by the concept of pseudospin whose direction simply reflects the relative phase $\phi_{AB}$ of two symmetric sublattices[7]. In the band of electrons in condensed matter, there can be a beautiful order of quantum phases dictated by crystalline symmetries like pseudospin chirality responsible for various unusual quantum phenomena[6-9].

A key idea to condensed-matter dark states is based on a double two-level quantum system, where there are two pairs of sublattices indexed A to D in Fig. 1a-d. If two pairs of sublattices are related by glide-mirror symmetries, their relative phases $\phi_{AB/CD}$, $\phi_{AC/BD}$, and $\phi_{AD/BC}$ are found fully polarised to either 0 (even parity +) or $\pi$ (odd parity –) in the whole Brillouin zone. Considering all the possible combinations of $\phi_{AB/CD}$, $\phi_{AC/BD}$, and $\phi_{AD/BC}$, we found there are only four kinds of pseudospin states indexed by $\phi_{AB/CD}\phi_{AC/BD}\phi_{AD/BC}$: 000, 0$\pi\pi$, $\pi$0$\pi$, and $\pi\pi$0 and colour-coded in Fig. 1a-d by blue, red, yellow, and green, respectively. The 000 states (blue pseudospin) in Fig. 1a can be detected in angle-resolved photoemission (ARPES) with $p$-polarised light by constructive interference. However, the other 0$\pi\pi$, $\pi$0$\pi$, and $\pi\pi$0 states (red, yellow, and green pseudospins) in Fig. 1b-d, where two of $\phi_{AB/CD}$, $\phi_{AC/BD}$, and $\phi_{AD/BC}$ are $\pi$, cannot be detected by ARPES at the light of any polarisation due to double destructive interference, which are, namely, condensed matter dark states.

## Observation of dark states in PdSe$_2$

As a prototypical model system to demonstrate the proposed idea of dark states, we chose palladium diselenides (PdSe$_2$) for its crystal structure shown in Fig. 1e,f. There is the square planar arrangement of one Pd atom at the centre and four corner-sharing Se atoms around. The two such structural units are slightly tilted about $z$ (Fig. 1e) and $y$ (Fig. 1f) axes clockwise or counterclockwise to form the two sublattices labelled A and B in the upper sublayer. The inversion counterpart of this sublayer with the two sublattices labelled C and D is stacked as illustrated in Fig. 1e,f, which in turn forms two pairs of sublattices that can be viewed as a double two-level quantum system. Interestingly, each pair of AB/CD, AC/BD, and AD/BC is connected in rotation by the reflection in $x$, $y$, and $z$ followed by a half translation along $y$, $z$, and $x$ (space group *Pbca*, No. 61, Extended Data Table 1). That is, in the primitive cell of PdSe$_2$, there are two pairs of sublattices related by multiple glide-mirror symmetries.

The momentum-space Brillouin zone of bulk PdSe$_2$ depicted only for its octant in Fig. 1g is cuboid-shaped with reciprocal lattices, $a^* = 2\pi/a$, $b^* = 2\pi/b$, and $c^* = 2\pi/c$ (Fig. 1e-g). Along the major high-symmetry axes, the dispersion of valence bands obtained by first principles calculations is present in Fig. 1h. From the valence band maximum (VBM) to −0.8 eV, there is a single dispersive band of the mainly Pd $4d_{z^2}$ character with its tip located at the $\Gamma$ point. Figure 2a shows the expected band structure of PdSe$_2$ along $k_x$ over multiple Brillouin zones indexed by $\Gamma_{hkl}$ for each $\Gamma$ point at $k_x = ha^*$, $k_y = kb^*$, and $k_z = lc^*$ ($h, k, l = 0, 1, \cdots$)[10]. However, our ARPES data taken with the $p$-polarised light of 87 eV corresponding to $k_z = \Gamma_6$ in Fig. 2b shows a remarkable deviation from this picture: The valence band of PdSe$_2$ centred at $\Gamma_{006}$ is clearly observed as expected, but we found no signature of those centred at $\Gamma_{\bar{1}06}$ and $\Gamma_{106}$ (grey dotted lines) over their whole Brillouin zones.

More interestingly, ARPES data in Fig. 2c taken at exactly the same experimental conditions but $s$-polarisation reveal that even the valence band centred at $\Gamma_{006}$ completely disappears as compared by energy distribution curves in Fig. 2d. We show again in Fig. 2e made independent of light polarisation ($p$-pol + $s$-pol) that valence bands centred at $\Gamma_{\bar{1}06}$ and $\Gamma_{106}$ are practically not observed in the $k_x$ direction. The same is true for $k_y$ (Fig. 2f), in which there is no signature of those centred at $\Gamma_{0\bar{1}6}$ and $\Gamma_{016}$ at any polarisations and scattering planes (see Extended Data Fig. 1 for the fuller set of data in $k_x$ and $k_y$). As summarised in a constant energy map in Fig. 3a and line profiles in Fig. 2g, the valence band of PdSe$_2$ appears at only one of the nine cuboidal Brillouin zones at the $k_z$ plane of $\Gamma_6$.

On the other hand, the $k_z$ dependence of ARPES data is shown in Fig. 2h. We found that the valence band centred at $\Gamma_{00l}$ alternatively vanishes at every odd $l$ number of Brillouin zones, regardless of light polarisation and scattering plane (see Extended Data Fig. 2 for the fuller set of data in $k_z$). Indeed, the constant energy map taken at $k_z = \Gamma_7$ in Fig. 3b show practically no signature of valence bands not only in $k_x$, but also in $k_y$. In the diagonal (or $s$) direction, however, we could clearly see those centred at $\Gamma_{\bar{1}\bar{1}7}$, $\Gamma_{1\bar{1}7}$, $\Gamma_{\bar{1}17}$, and $\Gamma_{117}$ (see Extended Data Fig. 2 for the fuller set of data in $k_s$). That is, the valence band of PdSe$_2$ appears at four of the nine cuboidal Brillouin zones for $k_z = \Gamma_7$ (see Extended Data Figs. 3 and 4 for the fuller set of Brillouin-zone and photon-energy dependence).

## Mechanism of dark states in PdSe$_2$

To explain our experimental findings, we paid our attention to the electronic wavefunctions of PdSe$_2$ obtained from the tight-binding model. Considering that valence bands consist of mainly four Pd sublattices (A–D)[11-13] that are symmetrical, the amplitudes of wavefunctions must be the same, but their relative phases of $\phi_{AB/CD}$, $\phi_{AC/BD}$, and $\phi_{AD/BC}$ may vary over the Brillouin zones, as shown in Fig. 3c-e. Interestingly, every cuboidal Brillouin zone of PdSe$_2$ is completely polarised to one of four kinds of pseudospin states: 000, 0$\pi\pi$, $\pi$0$\pi$, and $\pi\pi$0 states corresponding to blue, red, yellow, and green regions as defined in Fig. 3f. More importantly, no two adjacent Brillouin zones have the same colour, which shows the abrupt switching of relative phases across every zone boundary. This is the same mechanism of Dirac nodal lines formed along every zone boundary in the presence of multiple glide-mirror symmetries[14-16].

This momentum-space order of relative phases in a checkerboard pattern may affect ARPES, but this possibility has been largely overlooked so far.

ARPES intensity is the transition probability for optical excitations that can be approximated by Fermi's golden rule[17-19] to be proportional to the square of the matrix element $M^{\mathbf{k}}$ as

$$M^{\mathbf{k}} = \iiint \psi_f^\dagger (\mathbf{A} \cdot \mathbf{p}) \psi_i \, dxdydz, \quad (1)$$

where $\mathbf{A}$ is the electromagnetic vector potential, $\mathbf{p}$ is the electronic momentum operator, and $\psi_i$ and $\psi_f$ are initial-state and final-state wavefunctions, respectively. This $M^{\mathbf{k}}$ has been conventionally considered in terms of only the atomic orbital degree of freedom. However, unless there is only one atom of the same kind in the primitive cell, ARPES intensity should be dominated by quantum interference of sublattices no matter what their orbital character is as in graphite[20-22], graphene[23-25] and black phosphorus[26], although the disappearance of a whole band in the zone-by-zone manner with any light polarisations, scattering planes, and photon energies (the dark state) has not been observed in these cases.

As for PdSe$_2$, where each pair of sublattices is related in rotation by glide-mirror symmetries, $M^{\mathbf{k}}$ dominated by sublattice interference can be described in terms $\phi_{AB}$, $\phi_{AC}$, and $\phi_{AD}$ by

$$M^{\mathbf{k}} \approx (1 + e^{i\phi_{AB}} + e^{i\phi_{AC}} + e^{i\phi_{AD}}) \pm (1 + e^{i\phi_{AB}} + e^{i\phi_{AC}} + e^{i\phi_{AD}}), \quad (2)$$

Where ± means that the parity of two sublattice pairs related by glide-mirror symmetry with respect to a given scattering plane (see Extended Data Table 1 for parity pairs) is unchanged by *p*-polarised light but converted by *s*-polarised light. All the other summations come from integration over the scattering plane in equation (1), which stands for quantum interference between sublattices. This is valid for any orientation of scattering planes not only in *x* and *y*, but also in the other in-plane directions in between (see Methods for details).

In the case of *p*-polarisation, equation (2) becomes $M_p^{\mathbf{k}} \approx 1 + e^{i\phi_{AB}} + e^{i\phi_{AC}} + e^{i\phi_{AD}}$, and the $M_p^{\mathbf{k}}$ of the 000 states (blue in Fig. 1a) with $\phi_{AB} = \phi_{AC} = \phi_{AD} = 0$ is nonzero owing to constructive interference. However, that of $0\pi\pi$, $\pi 0\pi$, and $\pi\pi 0$ states (red, yellow, and green in Fig. 1b-d), where two of $\phi_{AB}$, $\phi_{AC}$, and $\phi_{AD}$ are $\pi$, becomes zero due to double destructive interference between two pairs of sublattices. In the case of *s*-polarisation, the $M_s^{\mathbf{k}}$ becomes zero for all four pseudospin states. If $M_p^{\mathbf{k}}$ and $M_s^{\mathbf{k}}$ are both zero, it is undetectable, because $M^{\mathbf{k}}$'s for the other circular polarisations must be zero as $M_p^{\mathbf{k}} \pm iM_s^{\mathbf{k}}$, where the ± sign stands for right and left circular polarisations. Therefore, the 000 states are bright only with *p*-polarised light, but $0\pi\pi$, $\pi 0\pi$, and $\pi\pi 0$ states cannot be detected with the light of any polarisations. Given that valence bands in PdSe$_2$ are completely polarised to one of $0\pi\pi$, $\pi 0\pi$, and $\pi\pi 0$ states for red, yellow, and green Brillouin zones in Fig. 3c-e, they correspond to dark states undetectable by any light polarisations, scattering planes, Brillouin zones, and photon energies.

With this model of $M^{\mathbf{k}}$ and initial-state wavefunctions taken from tight-binding calculations (Fig. 3c-e), ARPES simulations for the valence band of PdSe$_2$ centred at $k_z = \Gamma_{2n}$ and $k_z = \Gamma_{2n+1}$, where $n$ is integer, are present in Fig. 3g,h. They reproduce key aspects of our observations in that only one centred at $\Gamma_{00}$ in Fig. 3g and four centred at $\Gamma_{\bar{1}\bar{1}}$, $\Gamma_{1\bar{1}}$, $\Gamma_{\bar{1}1}$, and $\Gamma_{11}$ in Fig. 3h (corresponding to blue regions in Fig. 3c,d) appear as exactly observed in Fig. 3a,b. Not only that, but our ARPES simulations collectively reproduce the polarisation dependence of band dispersions and constant-energy maps, as summarised in Extended Data Fig. 5.

## Generalisation of dark states to other systems

In nature, there are surprisingly many materials with two pairs of sublattices connected by glide-mirror symmetries. One of the many such materials is cuprates, in the single-layer class (Bi$_2$Sr$_2$CuO$_{6+\delta}$ or Bi2201) of which there are four Cu sublattices labelled A to D in the primitive cell determined by X-ray diffraction (Fig. 4a,b)[27-29]. In the expected Fermi surface of cuprates (Fig. 4c,d), there are two bands labelled FS1 and FS2 from two sublayers (AB/CD), which are nearly degenerated by weak interlayer coupling. However, ARPES data of cuprates taken by *p*-polarised or *s*-polarised light have commonly shown the different periodicity as in Fig. 4f: There is little spectral intensity on those centred at $\Gamma_{00}$, $\Gamma_{\bar{1}\bar{1}}$, $\Gamma_{1\bar{1}}$, $\Gamma_{\bar{1}1}$, and $\Gamma_{11}$, which is called shadow bands[30,31]. A question here is why there is no strong intensity on shadow bands even in the presence of structure factors (clear Bragg peaks in diffraction[29]) arising from structural and/or antiferromagnetic origins. Identifying these hidden states is important to answer the long-standing question whether Fermi arcs[32-34] (related to pseudogap) or Fermi pockets[35-37].

The relative phase of sublattices taken by tight-binding calculations is plotted on the Fermi surface in Fig. 4c,d. Each Brillouin-zone segment of the large Fermi contours is polarised to one of the four kinds of pseudospin states (Fig. 3f) that are different between FS1 and FS2. Glide-mirror symmetries in *x*, *y*, and *z* of Bi2201 (space group *Pbnn*, No. 52, Extended Data Table 1) connect the sublattice pairs of AB/CD, AD/BC, and AB/CD, respectively. Because two of them relate the same sublattice pairs, the $M^{\mathbf{k}}$ of cuprates takes a slightly different form as $M^{\mathbf{k}} \approx (1 + e^{i\phi_{AB}}) \pm (e^{i\phi_{AC}} + e^{i\phi_{AD}})$ (see Methods for details). In the case of *p*-polarised light, $M^{\mathbf{k}}$ becomes the same as PdSe$_2$ and is nonzero only for 000 states (blue). However, for the case of *s*-polarisation $M^{\mathbf{k}}$ becomes $1 + e^{i\phi_{AB}} - e^{i\phi_{AC}} - e^{i\phi_{AD}}$, which is nonzero for $0\pi\pi$ states (red), but zero for $\pi 0\pi$ and $\pi\pi 0$ states (yellow and green) corresponding to dark states (Fig. 4e).

Based on this model of $M^{\mathbf{k}}$, ARPES simulations of the Fermi surface of cuprates are shown in Fig. 4g. Interestingly, blue (red) segments of FS1 and FS2 in Fig. 4c,d, which can be detected by *p*-polarised (*s*-polarised) light (Fig. 4e), join together to form the nearly identical Fermi contours centred at $\Gamma_{\bar{1}0}$, $\Gamma_{10}$, $\Gamma_{0\bar{1}}$, and $\Gamma_{01}$. On the other hand, those centred at $\Gamma_{00}$, $\Gamma_{\bar{1}\bar{1}}$, $\Gamma_{1\bar{1}}$, $\Gamma_{\bar{1}1}$, and $\Gamma_{11}$, which are the position of shadow bands in ARPES, consists of the yellow and green segments, that is, those in the position of shadow bands turn out to be the dark states. Therefore, even if any folded bands exist, they cannot be practically observed by ARPES, no matter what the origin of band folding is, unless the relative phase of sublattices in Fig. 4c,d is significantly changed. More importantly, if Coulomb repulsion $U$ is added to this model, ARPES simulations reproduce Fermi arcs as shown in Fig. 4h. Therefore, our findings explain not only the absence of strong ARPES intensity in the shadow bands of large Fermi surfaces, but also Fermi arcs of small Fermi pockets as half of each pocket is disappeared by the effect of dark states (Extended Data Fig. 6 for the doping and temperature dependence).

As can be seen in the case of shadow bands in cuprates, ARPES intensity patterns in violation of structural (or electronic) periodicity have been commonly observed. Another example of such materials is lead-halide perovskites, in the orthorhombic phase of which there are four Pb sublattices labelled A to D in the primitive cell (Fig. 5a)[38-40]. In the expected band structure for $k_z = \Gamma_{2n}$ and $k_z = \Gamma_{2n+1}$ in Fig. 5c,d, there are two valence bands labelled VB1 and VB2. However, ARPES data of $CsPbBr_3$ taken with the $p$-polarised light of 70 eV ($k_z = \Gamma_8$) in Fig. 5e and 60 eV ($k_z = \Gamma_7$) in Fig. 5f commonly show no signature of VB1 and VB2 centred at $\Gamma_{00}$. More importantly, for those centred at $\Gamma_{\bar{1}0}$ and $\Gamma_{10}$, we found only VB2 for $k_z = \Gamma_8$ and only VB1 for $k_z = \Gamma_7$. Even these two completely disappear for $s$-polarised light (Fig. 5g) regardless of $k_z$. As summarised in Fig. 5h, the VB1 at $k_z = \Gamma_7$ appears in four out of nine Brillouin zones, which looks like following the room-temperature cubic phase of $CsPbBr_3$ even though it was taken from the low-temperature orthorhombic phase, causing debates and confusions[41,42].

The relative phase of sublattices in $CsPbBr_3$ taken by tight-binding calculations is shown by colour in Fig. 5c,d. Each Brillouin-zone segment of valence bands is polarised to one of four kinds of pseudospin states (Fig. 3f), which are different for VB1 and VB2. Because the (glide) mirror symmetries of $CsPbBr_3$ (space group *Pnam*, No. 62, Extended Data Table 1) connects in rotation the sublattice pair of AD/BC, AB/CD, and AC/BD, the $M^{\mathbf{k}}$ of $CsPbBr_3$ is the same as that of $PdSe_2$, as summarised in Fig. 5b. Blue and red bands centred at $\Gamma_{\bar{1}0}$ and $\Gamma_{10}$ naturally explain why we can see only VB1 for $k_z = \Gamma_7$ and only VB2 for $k_z = \Gamma_8$. Yellow and green bands centred at $\Gamma_{00}$ also explain why there is no spectral intensity on VB1 and VB2, making ARPES intensity patterns as in Fig. 5h. Therefore, this is another natural consequence of dark states originating from the intrinsic quantum phases of initial state wavefunctions.

## Discussion

It has been universally observed in the study of ARPES that the periodicity of a system in its primitive cell does not match to that of ARPES intensity patterns. To remedy this mismatch, many have relied on unfolding based on the assumption that the strength of (translational) symmetry breaking is extremely weak[43,44]. It may be useful to deal with dopants, impurities, and imperfections, which are likely to be randomly distributed in actual samples but should be periodic in first principles. However, for the systems with structure factors coming from the molecular arrangement of atoms as in $PdSe_2$ and the octahedral distortion as in cuprates and $CsPbBr_3$, diffraction experiments show Bragg peaks of the super cell as sharp as those of the normal cell[10,29,39], indicating that the strength of symmetry breaking is not that weak. On the other hand, our model related to the photoemission structure factor[20-22] to come up with dark states arising from double destructive interference provides a natural explanation for this periodicity mismatch with no conflict to the presence of diffraction structure factors.

There is clear evidence that unfolding cannot fully account for our findings: The polarisation dependence of blue states in Fig. 2d cannot be explained by unfolding but can be naturally explained by our model of sublattice interference (Extended Data Fig. 5). More importantly, the valence band of PdSe$_2$ measured by *s*-polarised light overall disappears as in Fig. 2c, but if extremely narrow down the intensity scale, there is a remnant trace of valence bands with the characteristic intensity pattern as shown in Extended Data Fig. 7. This pattern cannot be explained by unfolding but can be exactly reproduced by our model as a small contribution of Se atoms located off the half-translation positions (with no violation of the space group). That is, once constituent sublattices deviate from half-translation positions, the polarisation of relative phases gets incomplete, and ARPES intensity disappears only along an axis as in graphite[20-22], graphene[23-25], and black phosphorus[26]. It is far more natural to explain not only simple materials like graphite, graphene and black phosphorus, but also complex materials like cuprates and perovskites by the unified picture of sublattice interference.

Our result is closely related to charge-density waves (CDW) and spin-density waves (SDW)[45] accompanied by translational symmetry breaking that forms a supercell with the *n* number of sublattices located close to 1/*n* fractional-translation positions. We would remind of the empirical facts in the field of ARPES that it has been hard to find a strong ARPES intensity on folded bands in CDW/SDW materials (e. g., 1*T*-TaS$_2$ and 1*T*-TiSe$_2$)[46,47] even at sufficiently low temperatures, where long-range CDW/SDW order is expected to be fully developed. This is weird in the picture to explain the absence of ARPES intensity in the backfolded bands with the lack of coherence. However, as more fully discussed in Extended Data Fig. 8, our model can naturally explain not only little ARPES intensity of folded bands as part of the dark states, but also gradually diminishing spectral intensity near back folding as a natural consequence of incomplete phase polarisations near the zone boundaries induced by symmetry breaking.

The multiple Coulomb wavelets become simplified into only four kinds (000, 0$\pi\pi$, $\pi$0$\pi$, $\pi\pi$0) in materials with two pairs of sublattices related by multiple glide-mirror symmetries. Then, the selection rule becomes as stringent as in atomic systems since the role of orbital angular momentum properties is replaced by the equally well-defined sublattice degree of freedom. That is, the mechanism of dark states comes from the intrinsic property of initial states that have nothing to do with final state effects and hence no photon-energy dependence. If the final state of $M^\mathbf{k}$ is replaced by a conduction state, it takes the same form as Equation (2) for the conduction state of even parity (only the ± sign will be switched for that of odd parity). This can be used to explain the strong linear dichroism of PdSe$_2$ in absorption[48] and second harmonic generation[49]. This hidden pseudospin polarisation also enables us to understand the band structure of lead halide perovskites[41,42] to explain their optoelectronic properties[50].

Our findings enable one to extract the complete map of initial-state quantum phases out of ARPES data that become closely analysable based on the model of sublattice interference, which would be helpful to determine the true strength of broken symmetry. More generally, the sublattice degree of freedom, which has been overlooked thus far, should be taken into consideration in the study of correlated phenomena and optoelectronic characteristics.

**Acknowledgements** This work was supported by the National Research Foundation (NRF) of Korea funded by the Ministry of Science and ICT (Grants No. NRF-2021R1A3B1077156, NRF-RS-2024-00416976, NRF-RS-2022-00143178 to K.S.K., NRF-2020R1I1A3073680 to K.L.), Yonsei Signature Research Cluster Program (2023-22-0004) to K.S.K., and Industry-Academy joint research program between Samsung Electronics and Yonsei University to Y.Y. This research used resources of the Advanced Light Source, which is the DOE Office of Science User Facility under the Contract No. DE-AC02-05CH11231. We acknowledge Diamond Light Source for time on Beamline I05 under Proposals SI30270 and SI35764.

**Author contributions** M.K. performed ARPES experiments on $PdSe_2$ with help from Y.C., C.J., E.R., and A.B. Y.C and Y.K. performed ARPES experiments on cuprates with help from T.K.K., C.C., C.J., E.R., and A.B. Y.K. performed ARPES experiments on $CsPbBr_3$ with help from S.C., Y.C., T.K.K., C.C., J.D., C.J., E.R., and A.B. J.W.P., J.P., Y.Y., D.S., J.H.R., and K.L. synthesised and provided samples. Y.C. carried out tight-binding band calculations and spectral simulations. K.S.K. conceived and supervised the project. Y.C., M.K., Y.K., and K.S.K wrote the manuscript with contributions from all other co-authors.

**Competing interests** The authors declare no competing interests.

## Additional information
**Extended data** for this paper is available at.

**Correspondence and requests for materials** should be addressed to Keun Su Kim.

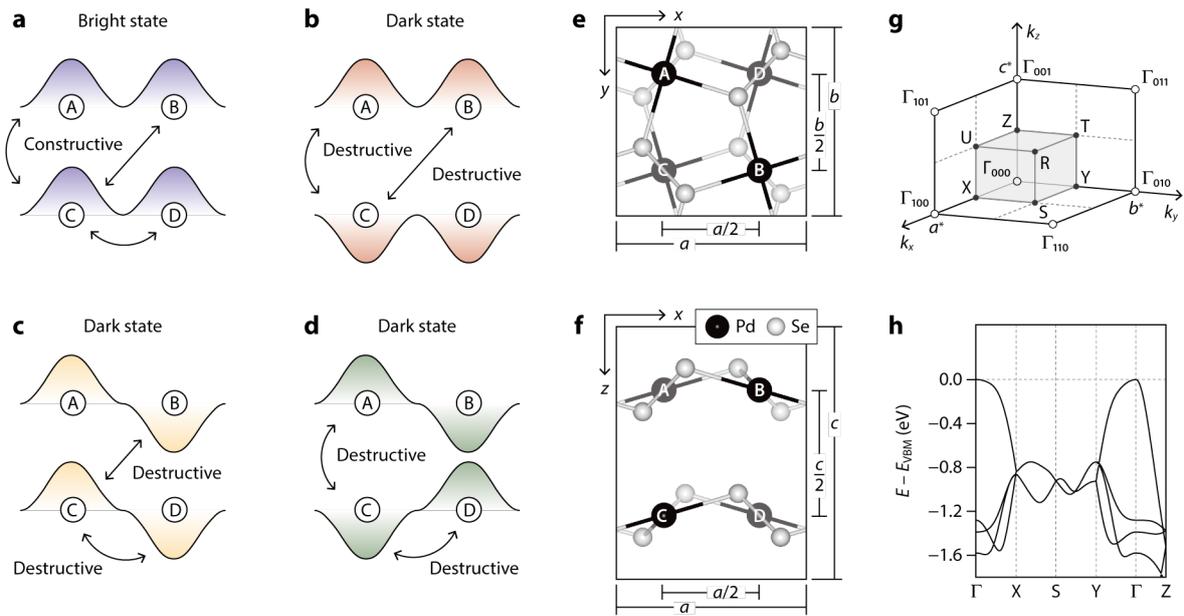

**Fig. 1 | Dark state of condensed matter and PdSe$_2$ as a material candidate. a-d,** Concept of bright and dark states in a system with two pairs of sublattices. Open circles labelled A to D represent sublattice atoms. The wavy lines in blue, red, yellow, and green illustrate four possible combinations of relative phases between sublattices (or pseudospins), indexed by $\phi_{AB/CD}\phi_{AC/BD}\phi_{AD/BC}$: 000 (**a**), $0\pi\pi$ (**b**), $\pi 0\pi$ (**c**), and $\pi\pi 0$ (**d**) states. The double-head arrows indicate constructive or destructive interference, leading to bright states in **a** and dark states in **b-d**. **e,f,** Crystal structure of PdSe$_2$ illustrated by a ball-and-stick model and viewed from the top (**e**) and side (**f**). The four Pd sublattices in the primitive cell are labelled from A to D. **g,** Brillouin zone of PdSe$_2$ shown in grey for an octant. The $\Gamma$ points marked by open circles are labelled as $\Gamma_{hkl}$ ($h$, $k$, $l$ = 0, 1, …). **h,** Valence bands of PdSe$_2$ obtained from first principles calculations and plotted along a few high-symmetry points marked by black dots in **g**.

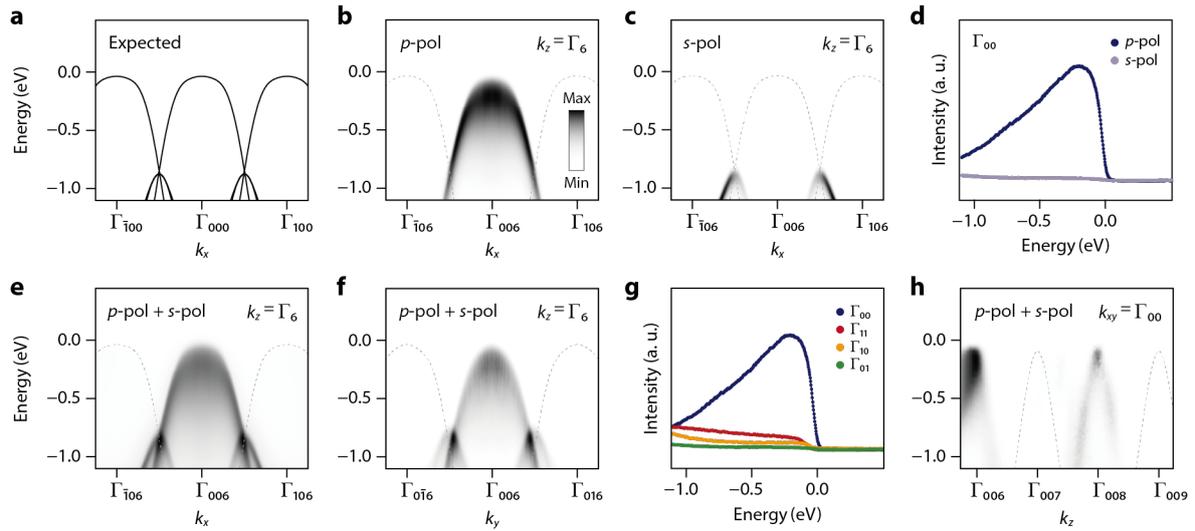

**Fig. 2 | Discovery of dark states in the band structure of PdSe$_2$. a,** Band structure of PdSe$_2$ in $k_x$, expected from first principles band calculations. **b,c,** ARPES data of PdSe$_2$ at 80 K, taken along $k_x$ with the *p*-polarised (**b**) and *s*-polarised (**c**) light of 87 eV corresponding to $k_z = \Gamma_6$. The same colour scale is applied to **b** and **c** normalised by the background. **d,** Line profiles taken at $\Gamma_{00}$ to compare data in **b** and **c**. **e,** ARPES data plotted by adding those in **b,c** (*p*-pol + *s*-pol), making it independent of light polarisation. **f,** ARPES data of PdSe$_2$ taken at $k_z = \Gamma_6$ with *p*-polarised and *s*-polarised light (*p*-pol + *s*-pol) along $k_y$. **g,** Line profiles taken at $\Gamma_{00}$, $\Gamma_{11}$, $\Gamma_{10}$, and $\Gamma_{01}$ to compare those of blue (000), red (0ππ), yellow (π0π), green (ππ0) states. **h**, ARPES data of PdSe$_2$ taken at $k_{xy} = \Gamma_{00}$ along $k_z$ with the photon energy of 80–200 eV. The inner potential is set at 10 eV. Grey dotted lines in **b-h** mark the expected location of bands.

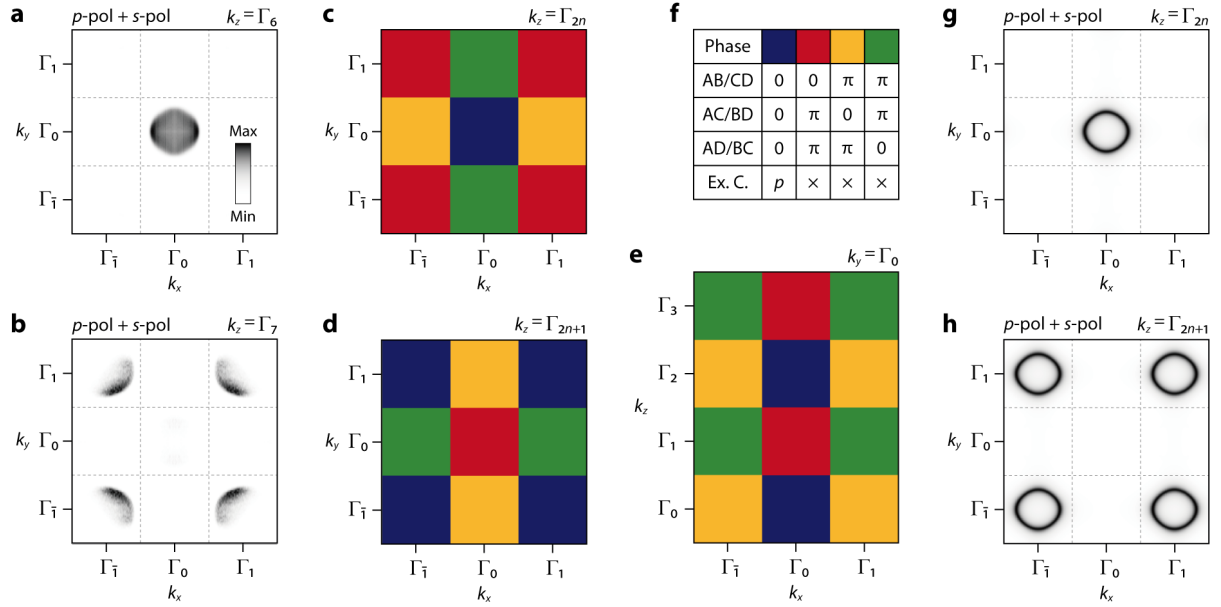

**Fig. 3 | Polarisation of relative phases between sublattices in PdSe$_2$. a,b,** ARPES data of PdSe$_2$ at 80 K, taken with *p*-polarised light and *s*-polarised light at $k_z = \Gamma_6$ (**a**) and $k_z = \Gamma_7$ (**b**) and plotted at $E = -0.4$ eV as a function of $k_x$ and $k_y$. Grey dotted lines show zone boundaries. **c-e,** Relative phases between sublattices (defined in **f**), calculated by tight-binding models and plotted for $k_z = \Gamma_{2n}$ (**c**) and $k_z = \Gamma_{2n+1}$ (**d**) as a function of $k_x$ and $k_y$, and for $k_y = \Gamma_0$ (**e**) as a function of $k_x$ and $k_z$. **f,** Four possible combinations of relative phases between sublattices ($\phi_{AB/CD}$, $\phi_{AC/BD}$, and $\phi_{AD/BC}$) colour-coded by blue, red, yellow, and green. The polarisation of light for excitations are shown in the bottom row, in which the cross marks represent dark states. **g,h,** Corresponding ARPES simulations based on the model of sublattice interference (see Methods) at $k_z = \Gamma_{2n}$ (**g**), and $k_z = \Gamma_{2n+1}$ (**h**).

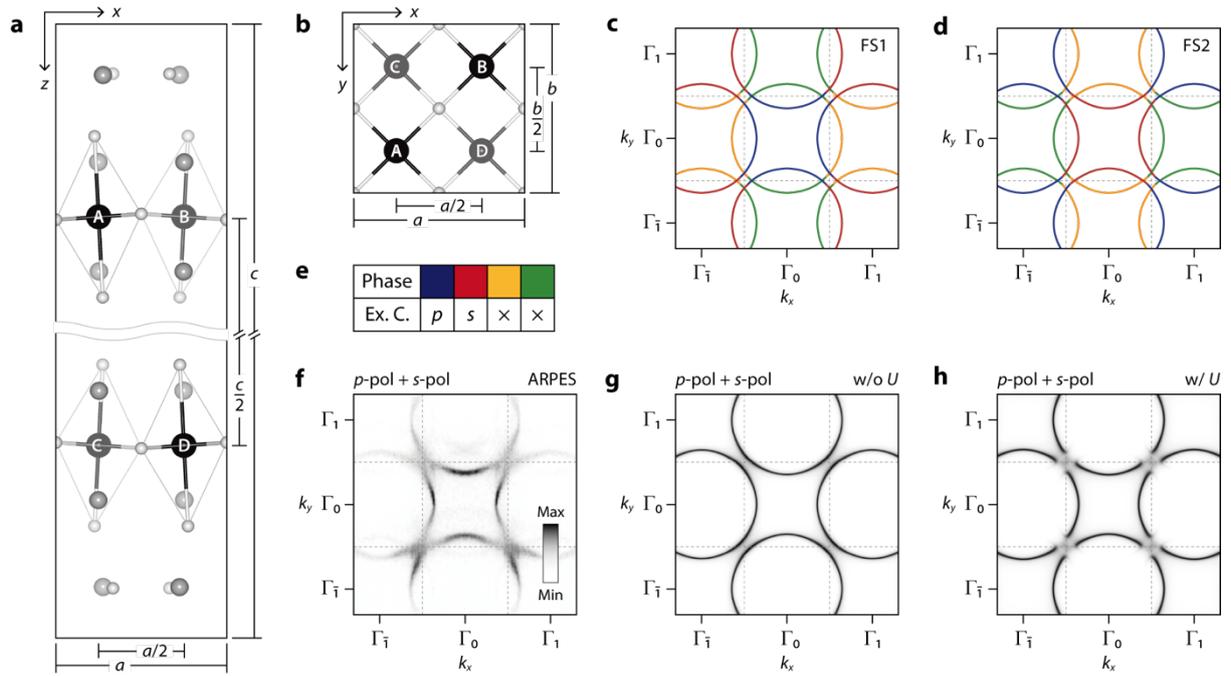

**Fig. 4 | Dark states in the band structure of cuprates. a,b,** Crystal structure of single-layer cuprates (or Bi2201) illustrated by the ball-and-stick model and viewed from the side (**a**) and the top (**b**). Grey rhombuses are $CuO_6$ octahedrons, and the 4 Cu sublattices in the primitive cell are labelled from A to D. **c,d,** Band structure of single-layer cuprates obtained by tight-binding calculations and shown separately for two nearly degenerate Fermi surfaces, FS1 (**c**) and FS2 (**d**). The colour of the Fermi surfaces means the relative phases between sublattices defined and colour-coded in Fig. 3f. **e,** Four kinds of the relative phases between sublattices (sublattice pseudospins) defined and colour-coded in Fig. 3f. The polarisation conditions for their excitations are shown in the bottom row, where the cross marks represent dark states. **f,** ARPES data of $Bi_{1.5}Pb_{0.5}Sr_2CaCu_2O_{8+\delta}$ ($T_c$ = 92 K), taken by $p$-polarised and $s$-polarised light and plotted at the Fermi energy as a function of $k_x$ and $k_y$. **g,h,** ARPES simulations based on the model of sublattice interference without Coulomb repulsion $U$ (**g**) and with $U$ (**h**).

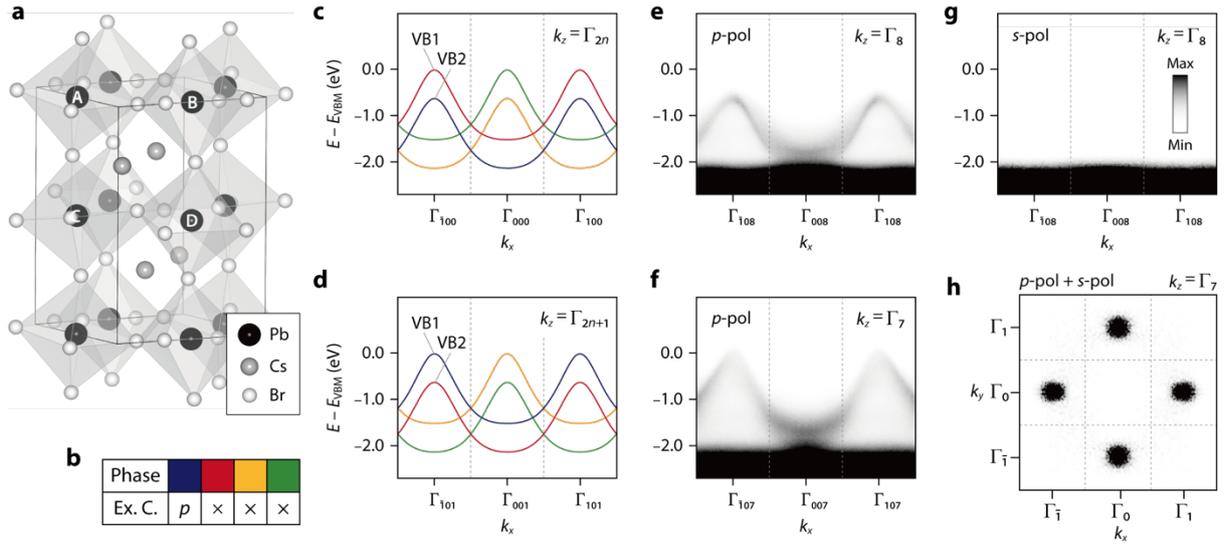

**Fig. 5 | Dark states in the band structure of orthorhombic CsPbBr$_3$. a,** Crystal structure of CsPbBr$_3$ in the orthorhombic phase. The shaded areas show tilted PbBr$_6$ octahedrons, and the 4 Pb sublattices in the primitive cell (black cuboid) are labelled from A to D. **b,** Four kinds of relative phases between sublattices (or pseudospins) defined and colour-coded in Fig. 3f. The polarisation of light for their excitations is shown in the bottom, where the cross marks represent dark states. **c,d,** Band structure of orthorhombic CsPbBr$_3$ taken by tight-binding calculations and plotted in $k_x$ at constant $k_z = \Gamma_{2n}$ (**c**) and $k_z = \Gamma_{2n+1}$ (**d**). The colour of VB1 and VB2 means the relative phases between sublattices, as defined and colour-coded in Fig. 3f. **e-g,** ARPES data of CsPbBr$_3$ at 100 K, taken in $k_x$ by $p$-polarised light of 70 eV corresponding to $k_z = \Gamma_8$ (**e**) and that of 60 eV corresponding to $k_z = \Gamma_7$ (**f**), and $s$-polarised light at $k_z = \Gamma_8$ (**g**). **h,** ARPES data of CsPbBr$_3$ taken at $k_z = \Gamma_7$ by $p$-polarised and $s$-polarised light and plotted at $E = E_{VBM} - 0.2$ eV as a function of $k_x$ and $k_y$. Grey dotted lines in **c-h** show zone boundaries.

## Methods

**ARPES experiments.** We have carried out ARPES experiments at Beamline 7.0.2 (MAESTRO) and Beamline 4.0.3 (MERLIN), Advanced Light Source, and at Beamline I05, Diamond Light Source. The microARPES end-stations are equipped with a hemispherical electron analyser with the 6-axis cryogenic manipulator. The high-flux photon beam is monochromatised in the range of 40–200 eV and focussed onto the surface of samples about 50 μm in diameter. The single crystals of $PdSe_2$, $Bi_{1.5}Pb_{0.5}Sr_2CaCu_2O_{8+\delta}$ ($T_c$ = 92 K) and $CsPbBr_3$ were *in situ* cleaved to prepare clean surfaces in the ultrahigh vacuum chamber whose base pressure was better than $5 \times 10^{-11}$ torr. It has been extremely difficult to have a uniform cleavage of $CsPbBr_3$ due to its three-dimensional character and beam damage. To overcome these, we have carefully scanned over the sample to secure a few best spots, where the sample surface happens to be locally uniform, and finished each measurement at a spot within 30 minutes. The sample temperature for $Bi_{1.5}Pb_{0.5}Sr_2CaCu_2O_{8+\delta}$ was set to 5–15 K, and that for $PdSe_2$ and $CsPbBr_3$ was set to 80–100 K to avoid charging effects. With these settings, energy and $k$ resolutions were better than 10 meV and 0.01 Å$^{-1}$. As for scattering geometry, the analyser slit in MERLIN and I05 is oriented vertically to the scattering plane with the variable light polarisation of linear horizontal (LH), linear vertical (LV), circular left (CL), and circular right (CR). In MAESTRO, the analyser slit orientation could be set to either vertical or horizontal owing to the rotatable analyser, which enables us to examine the possible combination of scattering planes (along $x$, $y$, or $s$), and linearly polarised light (LH or LV), as summarised in Extended Data Figs. 1–4.

**Tight-binding band calculations.** We constructed Slater-Koster tight-binding Hamiltonian for $PdSe_2$, Bi2201, Bi2212, and $CsPbBr_3$, considering the orbital symmetry of valence bands[51]. In the primitive cell of $PdSe_2$, there are 4 Pd atoms and 8 Se atoms (Fig. 1e,f), and their atomic coordinate is given by previous reports[52] (space group *Pbca*, No. 61, Extended Data Table 1). The valence band of $PdSe_2$ consists of mainly Pd $4d_{z^2}$ with a minor contribution from Se $4p_z$, and the Hamiltonian with the basis of only 4 Pd sublattices ($C_A$, $C_B$, $C_C$, $C_D$) can be written as

$$H_{Pd/Cu} = \begin{pmatrix} f_{AA} & f_{AB} & f_{AC} & f_{AD} \\ f_{AB} & f_{AA} & f_{AD} & f_{AC} \\ f_{AC} & f_{AD} & f_{AA} & f_{AB} \\ f_{AD} & f_{AC} & f_{AB} & f_{AA} \end{pmatrix}, (3)$$

where $f_{AA} = 2t_4\cos(ak_x) + 2t_5\cos(bk_y) + 2t_6\cos(ck_z)$, $f_{AB} = 4t_1\cos(0.5ak_x)\cos(0.5bk_y)$, $f_{AC} = 4t_3\cos(0.5bk_y)\cos(0.5ck_z)$, and $f_{AD} = 4t_2\cos(0.5ak_x)\cos(0.5ck_z)$. $t_i$ is the hopping parameter for the *i*th nearest-neighbour sites ($i$ = 1, 2, ⋯, 5), which is optimised to reproduce the experimental dispersion of valence bands in Fig. 2b-h and Extended Data Figs. 1 and 2. The relative phase between $i$ and $j$ sublattices is obtained from $C_i/C_j$ ($i, j$ = A, B, C, D), which takes the form of $\exp(i\phi_{ij})$. Considering that $\phi_{AB}$, $\phi_{AC}$, and $\phi_{AD}$ are polarised to either 0 or π, the four possible combinations are colour-coded as in Fig. 3f and plotted over Brillouin zones in Fig. 3c-e.

The relative phase patterns in Fig. 3c-e are unchanged even though we include Se atoms and Pd-Se coupling in the model. We constructed the complete 12-band Hamiltonian with the basis of 4 Pd atoms and 8 Se atoms as

$$H_{PdSe_2} = \begin{pmatrix} H_{Pd} & H_{Pd\text{-}Se} \\ H_{Pd\text{-}Se} & H_{Se} \end{pmatrix},$$

where $H_{Se}$ is the 8 × 8 tight-binding Hamiltonian that includes up to 12 hopping parameters between Se atoms. $H_{Pd\text{-}Se}$ is the Hamiltonian that includes the 3 nearest-neighbour hopping parameters between Pd and Se atoms. Even with this Pd-Se coupling, the relative phases of Pd sublattices remain the same as Fig. 3c-e, as protected by the crystal symmetries of $PdSe_2$. On the other hand, those of Se atoms are plotted in Extended Data Fig. 7, which are different from Fig. 3c-e because they are located off the half-translation positions.

Conventionally, the tight-binding model of cuprates has been constructed with only one Cu atom per unit cell[53], which is strictly speaking not the primitive cell. We thus extended it to the more complete model with $4n$ Cu sublattices in the primitive cell of $n$-layer cuprates, and their atomic coordinate is given by past reports[27,28] (space group *Pbnn*, No. 52, Extended Data Table 1). The tight-binding Hamiltonian of single-layer cuprates (Bi2201) with the basis of 4 Cu sublattices ($C_A$, $C_B$, $C_C$, $C_D$) takes the same form as equation (3), where $f_{AA} = 2t_2[\cos(ak_x) + \cos(bk_y)] + 4t_3\cos(ak_x)\cos(bk_y)$, $f_{AB} = 2t_1\cos(0.5ak_x)\cos(0.5bk_y) + 4t_4[\cos(1.5ak_x)\cos(0.5bk_y) + \cos(0.5ak_x)\cos(1.5bk_y)]$, $f_{AC} = 2t_5\cos(0.5bk_y)\cos(0.5ck_z)[2\sin(0.5ak_x)\sin(0.5bk_y)]^2$, and $f_{AD} = 2t_5\cos(0.5ak_x)\cos(0.5ck_z)[2\sin(0.5ak_x)\sin(0.5bk_y)]^2$. $t_i$ is the tight-binding hopping parameter for the $i$th nearest-neighbour Cu sites ($i = 1, 2, \cdots, 5$), optimised to reproduce the experimental Fermi contours in Fig. 4f. The relative phase between $i$ and $j$ sublattices is obtained from $C_i/C_j$ ($i, j$ = A, B, C, D), which also takes the form of $\exp(i\phi_{ij})$. Considering that $\phi_{AB}$, $\phi_{AC}$, and $\phi_{AD}$ are polarised to either 0 or π, four possible combinations are colour-coded as defined in Fig. 3f, and shown along two nearly degenerate Fermi contours labelled FS1 and FS2 in Fig. 4c,d.

The relative phase patterns in Fig. 4c,d are unchanged even for bilayer cuprates (Bi2212) in the model. We constructed the tight-binding Hamiltonian for Bi2212 with the basis of 8 Cu sublattices ($C_{A1}$, $C_{B1}$, $C_{C1}$, $C_{D1}$, $C_{A2}$, $C_{B2}$, $C_{C2}$, $C_{D2}$) and an additional hopping parameter $t_6$ that describes coupling between each two layers. Even with this coupling, the relative phases of Cu sublattices ($\phi_{A1B1/A2B2}$, $\phi_{A1C1/A2C2}$, $\phi_{A1D1/A2D2}$) remain the same as Fig. 4c,d, as protected by the crystal symmetries of Bi2212 or Pb-doped Bi2212 (see Extended Data Table 1)[54].

To consider the effect of Coulomb repulsion, we constructed the Hubbard Hamiltonian[55] with the basis of 4 Cu sublattices and electron spins ($C_A\uparrow$, $C_B\uparrow$, $C_C\uparrow$, $C_D\uparrow$, $C_A\downarrow$, $C_B\downarrow$, $C_C\downarrow$, $C_D\downarrow$) as

$$H_H = \begin{pmatrix} H_{Cu} + H_U & 0 \\ 0 & H_{Cu} - H_U \end{pmatrix}, H_U = \begin{pmatrix} U & 0 & 0 & 0 \\ 0 & -U & 0 & 0 \\ 0 & 0 & U & 0 \\ 0 & 0 & 0 & -U \end{pmatrix},$$

where $U$ is the on-site Coulomb repulsion set to $0.2t_1$ for ARPES simulations in Fig. 4h.

As for the orthorhombic phase of CsPbBr$_3$, there are 4 Pb atoms in the primitive cell, and their atomic coordinate is given by previous reports[56] (space group *Pnam*, No. 62, Extended Data Table 1). The tight-binding Hamiltonian with the basis of 4 Pb sublattices ($C_A$, $C_B$, $C_C$, $C_D$) without considering spin-orbit coupling for valence bands of mainly Pb 6*s* can be written as

$$H_{Pb} = \begin{pmatrix} H_{AA} & H_{AB} & H_{AC} & H_{AD} \\ H_{AB}^\dagger & H_{BB} & H_{BC} & H_{BD} \\ H_{AC}^\dagger & H_{BC}^\dagger & H_{CC} & H_{CD} \\ H_{AD}^\dagger & H_{BD}^\dagger & H_{DC}^\dagger & H_{DD} \end{pmatrix},$$

where $H_{ii}$ and $H_{ij}$ are the Hamiltonian for coupling between *i* sublattices and between *i* and *j* sublattices (*i*, *j* = A, B, C, D). They can be written with the basis of 4 orbitals (*s*, $p_x$, $p_y$, $p_z$) as

$$H_{ii} = \begin{pmatrix} E^s & 0 & 0 & 0 \\ 0 & E^p & 0 & 0 \\ 0 & 0 & E^p & 0 \\ 0 & 0 & 0 & E^p \end{pmatrix}, H_{AB/CD} = \begin{pmatrix} g_0 & g_1 & g_2 & 0 \\ g_1^\dagger & g_3 & g_5 & 0 \\ g_2^\dagger & g_5^\dagger & g_4 & 0 \\ 0 & 0 & 0 & g_6 \end{pmatrix}, H_{AC/BD} = \begin{pmatrix} f_4 & 0 & 0 & f_5 \\ 0 & f_6 & 0 & 0 \\ 0 & 0 & f_6 & 0 \\ f_5^\dagger & 0 & 0 & f_7 \end{pmatrix},$$

and $H_{AD/BC} = 0$, where $g_0 = 4t_{AB}^{ss}\cos(0.5ak_x)\cos(0.5bk_y)$, $g_1 = 4i\cos(\theta_1)t_{AB}^{sp}\sin(0.5ak_x)\cos(0.5bk_y)$, $g_2 = 4i\cos(\theta_2)t_{AB}^{sp}\cos(0.5ak_x)\sin(0.5bk_y)$, $g_3 = 4f(\theta_1)\cos(0.5ak_x)\cos(0.5bk_y)$, $g_4 = 4f(\theta_2)\cos(0.5ak_x)\cos(0.5bk_y)$, $g_5 = 4f(\theta_{12})\sin(0.5ak_x)\sin(0.5bk_y)$, $g_6 = 4t_{AB}^{pp\pi}\cos(0.5ak_x)\cos(0.5bk_y)$, $f(\theta_1) = t_{AB}^{pp\sigma}\cos^2(\theta_1) + t_{AB}^{pp\pi}\sin^2(\theta_1)$, $f(\theta_2) = t_{AB}^{pp\sigma}\cos^2(\theta_2) + t_{AB}^{pp\pi}\sin^2(\theta_2)$, $f(\theta_{12}) = (t_{AB}^{pp\sigma} - t_{AB}^{pp\pi})\cos(\theta_1)\cos(\theta_2)$, $f_4 = 2t_{AC}^{ss}\cos(0.5ck_z)$, $f_5 = 2it_{AC}^{sp}\sin(0.5ck_z)$, $f_6 = 2t_{AC}^{pp\pi}\cos(0.5ck_z)$, and $f_7 = 2t_{AC}^{pp\sigma}\cos(0.5ck_z)$. $\theta_1$ and $\theta_2$ are related to a slight difference in lattice parameters, as defined in Fig. 15 of ref. [57]. $E^m$ is the on-site energy for *m* orbitals, and $t_{ij}^{mn}$ is the hopping parameter between *m* orbital at *i* site and *n* orbital at *j* site (*m*, *n* = *s*, *p*). $E^m$ and $t_{ij}^{mn}$ are optimised to reproduce the experimental dispersion of two valence bands labelled VB1 and VB2 in Fig. 5e,f. The relative phase of VB1 and VB2 between *i* and *j* sublattices is taken by $C_i/C_j$ (*i*, *j* = A, B, C, D), which is in the form of $\exp(i\phi_{ij})$. Considering that $\phi_{AB}$, $\phi_{AC}$, and $\phi_{AD}$ are polarised to either 0 or π, the four possible combinations are colour-coded as in Fig. 3f and plotted along VB1 and VB2 in Fig. 5c,d. We checked that the relative phase patterns in Fig. 5c,d are unchanged even if we include spin-orbit coupling in the model[57], as protected by the crystal symmetries of CsPbBr$_3$.

**Sublattice interference model.** We consider simple orthorhombic lattice systems with two pairs of sublattices ($C_A$, $C_B$, $C_C$, $C_D$) located at half-translation positions and related by multiple glide-mirror symmetries. Given that the bands of interest mainly consist of the 4 sublattices, the integrals in equation (1) can be approximated to the summation of coefficients of initial state wavefunctions. Then, $M^\mathbf{k}$ in the scattering plane of *xz* and *yz* can be written as

$$M_x^\mathbf{k} \approx C_I \left(1 \pm \frac{C_{M_y}}{C_I}\right)\left(1 + \frac{C_{M_x}}{C_I}\right)\left(1 + \frac{C_{M_z}}{C_I}\right),$$

$$M_y^{\mathbf{k}} \approx C_I \left(1 \pm \frac{C_{M_x}}{C_I}\right)\left(1 + \frac{C_{M_y}}{C_I}\right)\left(1 + \frac{C_{M_z}}{C_I}\right),$$

where $C_{M_i}$ is the sublattice counterpart connected by the (glide) mirror symmetry operation in the $i$ direction ($i = x, y, z$), and $C_I$ is the identity. The first term comes from the $\mathbf{A} \cdot \mathbf{p}$ term followed by integrating over the direction normal to the scattering plane. The $\pm$ sign means that the parity in this pair of sublattices is unchanged with LH or $p$-polarised light (+ parity) but converted with LV or $s$-polarised light (− parity) as in the case of graphene ($C_A \pm C_B$)[23-25]. The other two terms come from integration over the scattering plane of $xz$ or $yz$.

For orthorhombic systems, we have $M_i M_j = C_{2k}$ ($i, j, k = x, y, z$ and $i \neq j \neq k$) and $M_x M_y M_z = P$, where $C_{2k}$ is the twofold (screw) rotational symmetry in the $k$ direction, and $P$ is the inversion symmetry. With these relations, $M^{\mathbf{k}}$ in the scattering plane of $xz$ and $yz$ can be rewritten as

$$M_x^{\mathbf{k}} \approx \left(C_I + C_{M_x} + C_{M_z} + C_{C_{2y}}\right) \pm \left(C_{M_y} + C_{C_{2z}} + C_{C_{2x}} + C_P\right),$$
$$M_y^{\mathbf{k}} \approx \left(C_I + C_{M_y} + C_{M_z} + C_{C_{2x}}\right) \pm \left(C_{M_x} + C_{C_{2z}} + C_{C_{2y}} + C_P\right).$$

For PdSe$_2$ and CsPbBr$_3$, the sublattice pairs related by crystal symmetries are summarised in Extended Data Table 1. Putting them into the above equations, we obtain $M_x^{\mathbf{k}}$ and $M_y^{\mathbf{k}}$ as

$$M_x^{\mathbf{k}} \approx (C_A + C_B + C_C + C_D) \pm (C_A + C_B + C_C + C_D),$$
$$M_y^{\mathbf{k}} \approx (C_A + C_B + C_C + C_D) \pm (C_A + C_B + C_C + C_D).$$

Interestingly, we found $M_x^{\mathbf{k}}$ and $M_y^{\mathbf{k}}$ exactly the same. Since $M^{\mathbf{k}}$ in the scattering plane of any orientation in between $x$ and $y$ can be described by trigonometric projections to $M_x^{\mathbf{k}}$ and $M_y^{\mathbf{k}}$, $M^{\mathbf{k}}$ in this model should be independent of scattering planes. Dividing the above equations by $C_A$, one can obtain equation (2) in terms of $\phi_{AB}$, $\phi_{AC}$, and $\phi_{AD}$. As for the Se contributions, there are 8 Se atoms ($C_{A1}$, $C_{B1}$, $C_{C1}$, $C_{D1}$, $C_{A2}$, $C_{B2}$, $C_{C2}$, $C_{D2}$) in PdSe$_2$, and sublattice pairs related by their crystal symmetries yield $M_x^{\mathbf{k}} \approx (C_{A1} + C_{B1} + C_{C2} + C_{D2}) \pm (C_{C1} + C_{D1} + C_{A2} + C_{B2})$ and $M_y^{\mathbf{k}} \approx (C_{A1} + C_{C1} + C_{A2} + C_{C2}) \pm (C_{B1} + C_{D1} + C_{B2} + C_{D2})$.

As for single-layer cuprates or Pb-doped Bi2201, the key difference in the sublattice pairs related by crystal symmetries is that the two of the mirror symmetries in $x$ and $z$ connect the same sublattice pairs (Extended Data Table 1). This leads to a slightly different form of $M^{\mathbf{k}}$ as

$$M^{\mathbf{k}} \approx (C_A + C_B + C_B + C_A) \pm (C_D + C_C + C_C + C_D) \approx (C_A + C_B) \pm (C_C + C_D).$$

This naturally explains the difference between Fig. 3f and Fig. 4e that the excitation of $0\pi\pi$ states (red) by $s$-polarised light is forbidden for PdSe$_2$ and CsPbBr$_3$ but allowed for cuprates. This $M^{\mathbf{k}}$ of sublattice interference is basically unchanged by increasing the number of CuO$_2$ layers in the primitive cell. For example, bilayer cuprates or Bi2212 contains 8 Cu sublattices ($C_{A1}$, $C_{B1}$, $C_{C1}$, $C_{D1}$, $C_{A2}$, $C_{B2}$, $C_{C2}$, $C_{D2}$). Since its crystal symmetries are exactly the same as Bi2201, $M^\mathbf{k}$ is written in nearly the same form as ($C_{A1} + C_{B1} + C_{A2} + C_{B2}$) ± ($C_{C1} + C_{D1} + C_{C2} + C_{D2}$).

**Spectral simulations.** ARPES intensity $I(\mathbf{k}, E)$ is simulated based on the standard model[17-19] as a product of the square of $M^\mathbf{k}$ and the spectral function in the Lorentzian form as

$$I(\mathbf{k}, E) \approx \frac{\sigma}{(E - E^\mathbf{k})^2 + \sigma^2} |M^\mathbf{k}|^2,$$

where $E^\mathbf{k}$ is the band dispersion taken from tight-binding band calculations, $\sigma$ is the spectral width set to 0.03–0.08 eV. $M^\mathbf{k}$ is taken from the model of sublattice interference. The ARPES simulations made in this way are shown in Fig. 3, Fig. 4, and Extended Data Figs. 5–8.

## Data availability
All data are available from the corresponding author upon reasonable request. Source data are provided with this paper.

## Method-only references

**Extended Data Table 1 | Crystal symmetries and sublattice pairs.** The crystal symmetries of PdSe$_2$, Pb-doped Bi2201 / Bi2212, and CsPbBr$_3$ are listed. $M_i$ ($\widetilde{M}_i$) is (glide) mirror symmetry in the $i$ direction ($i = x, y, z$), and $C_{2i}$ ($\widetilde{C}_{2i}$) is twofold (screw) rotation symmetry in the $i$ direction. $P$ is the inversion symmetry. The half-translations of $\widetilde{M}_i$ and $\widetilde{C}_{2i}$ are shown in the sequence of $x, y, z$ in units of $a, b, c$, respectively. The sublattice pairs connected by corresponding crystal symmetry operations are shown in the form of $jk$ / $lm$, where $j, k, l, m$ = A, B, C, D.

| Material | Crystal symmetry | Half translation | Sublattice pair |
|---|---|---|---|
| PdSe$_2$ (*Pbca*, 61) | $\widetilde{M}_x$ | 0, ½, 0 | AB / CD |
| | $\widetilde{M}_y$ | 0, 0, ½ | AC / BD |
| | $\widetilde{M}_z$ | ½, 0, 0 | AD / BC |
| | $\widetilde{C}_{2x}$ | ½, 0, 0 | AB / CD |
| | $\widetilde{C}_{2y}$ | 0, ½, 0 | AC / BD |
| | $\widetilde{C}_{2z}$ | 0, 0, ½ | AD / BC |
| | $P$ | 0, 0, 0 | Identity |
| Bi2201 (*Pbnn*, 52) | $\widetilde{M}_x$ | 0, ½, 0 | AB / CD |
| | $\widetilde{M}_y$ | ½, 0, ½ | AD / BC |
| | $\widetilde{M}_z$ | ½, ½, 0 | AB / CD |
| | $C_{2x}$ | 0, 0, 0 | AC / BD |
| | $C_{2y}$ | 0, 0, 0 | Identity |
| | $\widetilde{C}_{2z}$ | 0, 0, ½ | AC / BD |
| | $P$ | 0, 0, 0 | AD / BC |
| CsPbBr$_3$ (*Pnam*, 62) | $\widetilde{M}_x$ | 0, ½, ½ | AD / BC |
| | $\widetilde{M}_y$ | ½, 0, 0 | AB / CD |
| | $M_z$ | 0, 0, 0 | AC / BD |
| | $\widetilde{C}_{2x}$ | ½, 0, 0 | AD / BC |
| | $\widetilde{C}_{2y}$ | 0, ½, 0 | AB / CD |
| | $\widetilde{C}_{2z}$ | 0, 0, ½ | AC / BD |
| | $P$ | 0, 0, 0 | Identity |

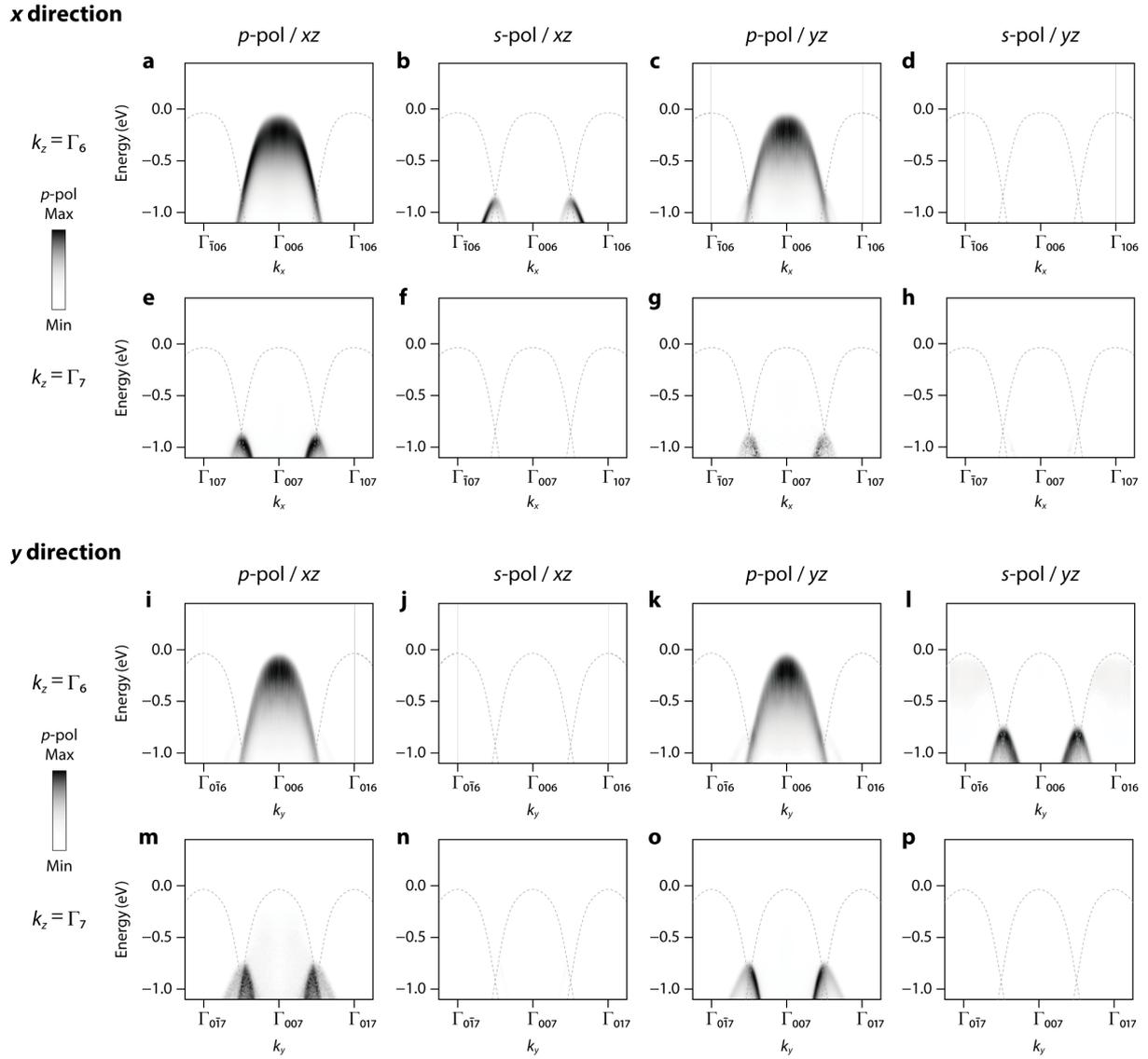

**Extended Data Fig. 2 | ARPES data of PdSe$_2$ in $k_x$ and $k_y$. a-p,** Complete set of ARPES data taken at 80 K along $k_x$ (**a-h**) and $k_y$ (**i-p**). The light polarisation and scattering plane are shown by *p*-pol / *xz* (**a,e,i,m**), *s*-pol / *xz* (**b,f,j,n**), *p*-pol / *yz* (**c,g,k,o**), and *s*-pol / *yz* (**d,h,l,p**), as marked on the top. The photon energy is 87 eV for $k_z = \Gamma_6$ (**a-d,i-l**) and 119 eV for $k_z = \Gamma_7$ (**e-h,m-p**), as marked on the left. After background normalisations, the same colour scale is applied to each pair of data taken by *p*-polarised light and *s*-polarised light. Grey dotted lines indicate the expected location of valence bands. It can be identified that the valence bands of PdSe$_2$ centred at all $\Gamma$ points but $\Gamma_{006}$ are dark states, regardless of scattering planes.

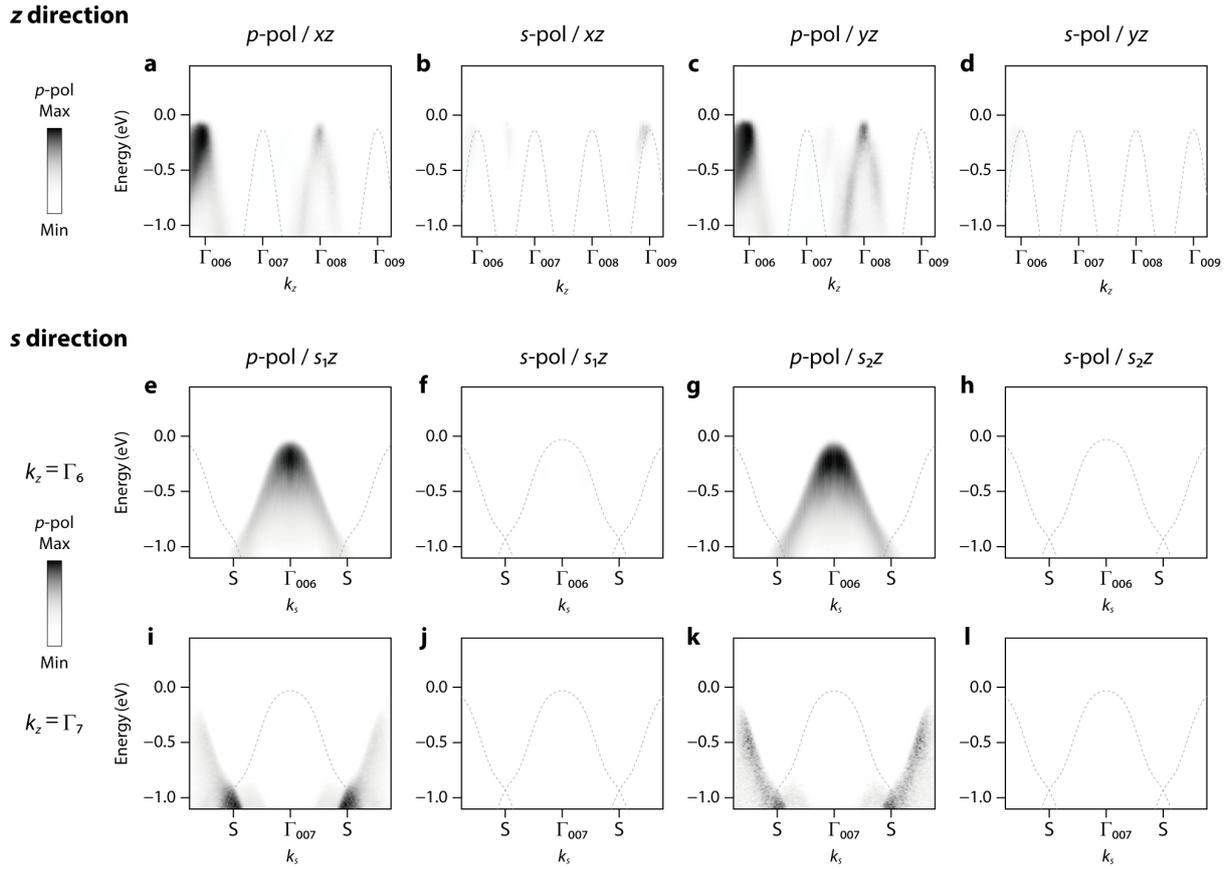

**Extended Data Fig. 3 | ARPES data of PdSe$_2$ in $k_z$ and $k_s$. a-d,** Complete set of ARPES data taken at 80 K along $k_z$. The light polarisation and scattering plane are shown by $p$-pol / $xz$ (**a**), $s$-pol / $xz$ (**b**), $p$-pol / $yz$ (**c**), and $s$-pol / $yz$ (**d**), as marked on the top. **e-l,** Complete set of ARPES data taken along $k_s$. The light polarisation and scattering plane are shown by $p$-pol / $s_1z$ (**e,i**), $s$-pol / $s_1z$ (**f,j**), $p$-pol / $s_2z$ (**g,k**), and $s$-pol / $s_2z$ (**h,l**), where $s_1$ and $s_2$ are two diagonal directions, as shown on the top. The photon energy is 87 eV for $k_z = \Gamma_6$ (**e-h**) and 119 eV for $k_z = \Gamma_7$ (**i-l**), as shown on the left. After background normalisations, the same colour scale is applied to each pair of data taken with $p$-polarised and $s$-polarised light. Grey dotted lines indicate the expected location of valence bands. It could be identified that the valence bands of PdSe$_2$ centred at $\Gamma_{007}$ and $\Gamma_{009}$ are dark states, regardless of scattering planes.

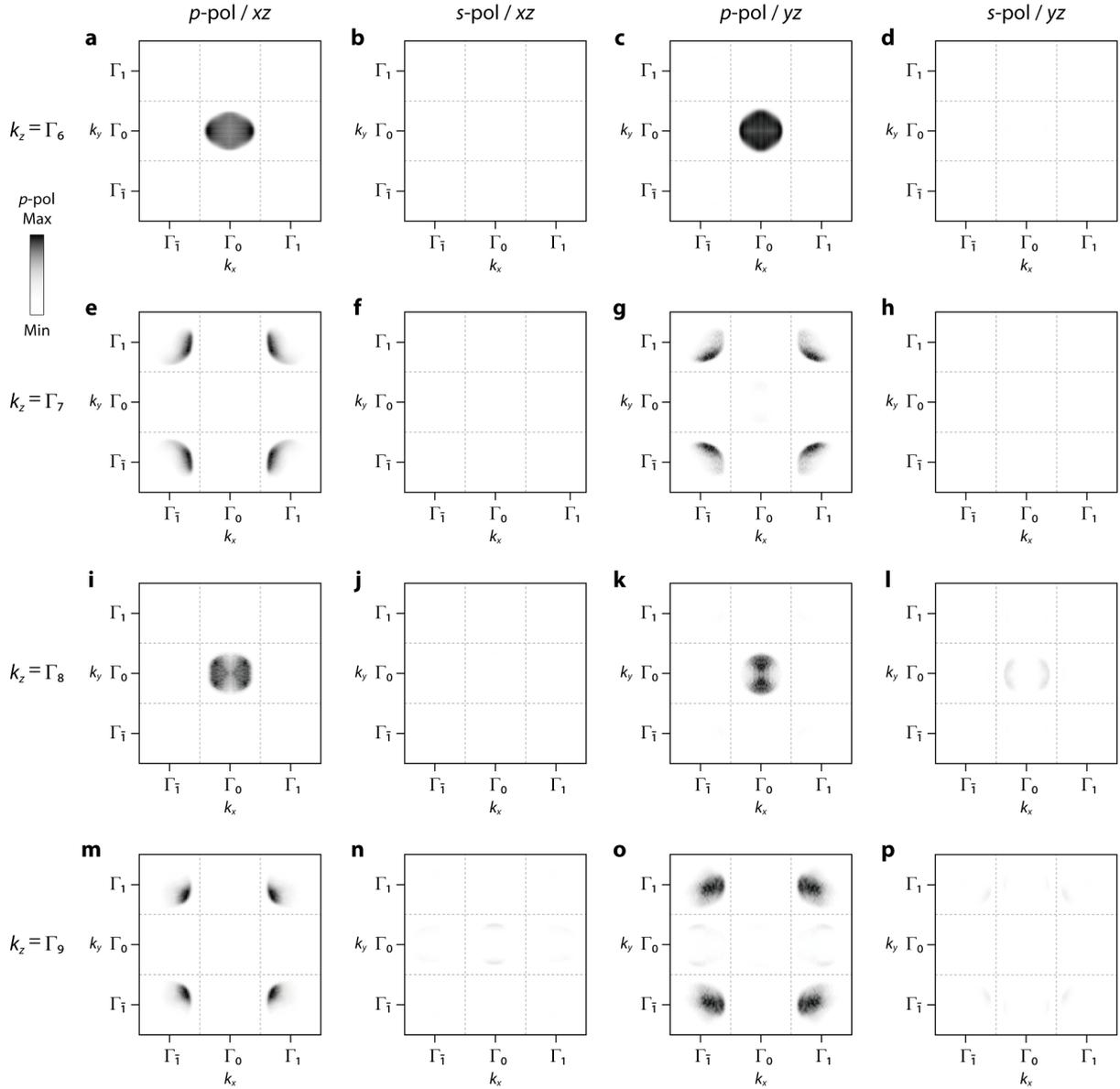

**Extended Data Fig. 4 | ARPES data of PdSe$_2$ in constant energy maps. a-p,** ARPES data taken at 80 K and plotted at $E = -0.4$ eV as a function of $k_x$ and $k_y$. The light polarisation and scattering plane are indicated by *p*-pol / *xz* (**a,e,i,m**), *s*-pol / *xz* (**b,f,j,n**), *p*-pol / *yz* (**c,g,k,o**), and *s*-pol / *yz* (**d,h,l,p**), as marked on the top. The photon energy is 87 eV for $k_z = \Gamma_6$ (**a-d**), 119 eV for $k_z = \Gamma_7$ (**e-h**), 154 eV for $k_z = \Gamma_8$ (**i-l**), and 197 eV for $k_z = \Gamma_9$ (**m-p**), as shown on the left. After background normalisations, the same colour scale is applied to each pair of data taken with *p*-polarised light and *s*-polarised light. Grey dotted lines show zone boundaries. We found that the valence bands of PdSe$_2$ in 13 out of the 18 Brillouin zones are dark states, regardless of scattering planes. The remnant intensity seen in **l,n** are due to Se contributions, as more fully discussed in Extended Data Fig. 8.

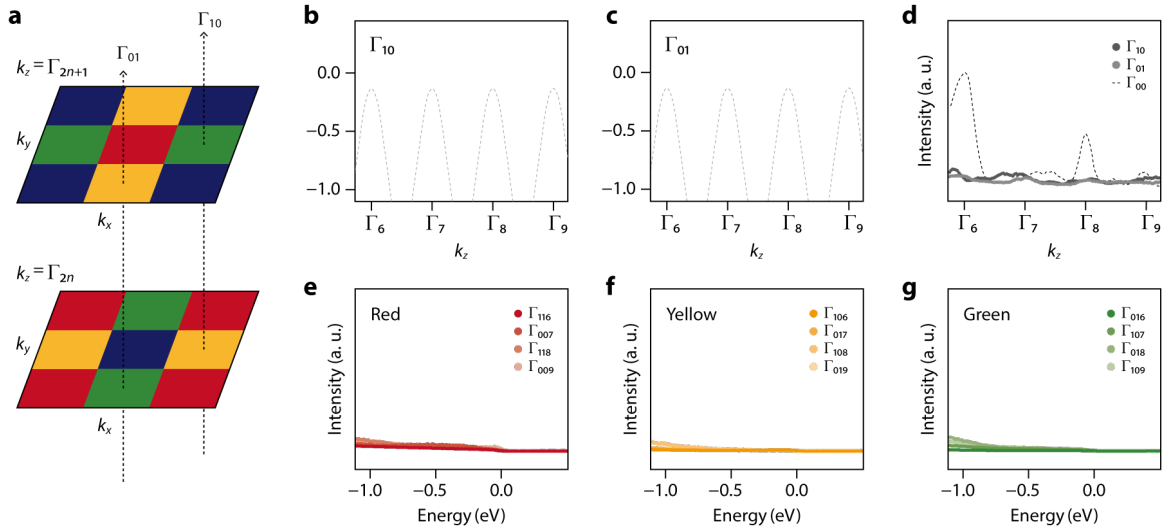

**Extended Data Fig. 5 | Photon energy dependence of dark states. a,** Alternating relative phases of PdSe$_2$ between $k_z = \Gamma_{2n}$ and $\Gamma_{2n+1}$. **b,c,** Photon-energy dependence of ARPES data taken at $k_{xy} = \Gamma_{10}$ (**b**) and $\Gamma_{01}$ (**c**), as indicated by dotted arrows in **a** (*p*-pol + *s*-pol, *xz* scattering plane). The colour scale is normalised based on the background with respect to those taken at $k_{xy} = \Gamma_{00}$ (Extended Data Fig. 3a,b), as shown in **d**. **d,** Momentum distribution curves (MDCs) taken at $E = -0.1$ eV along $k_z$ (80–200 eV) to make a comparison for those at $\Gamma_{10}$, $\Gamma_{01}$, and $\Gamma_{00}$. **e-g,** EDCs taken at four different $\Gamma_{hkl}$ points collected to separately show the photon-energy dependence of red (**e**), yellow (**f**), and green (**g**) states corresponding to dark states. As seen in **d-g**, we find little photon-energy dependence of dark states for all photon energies used here. This is because multiple Coulomb wavelets become simplified (or polarised) into only four kinds in materials with two pairs of sublattices connected by glide-mirror symmetries. Then, the selection rule becomes as stringent as in atomic systems since the role of orbital angular momentum properties is replaced by the equally well-defined sublattice degree of freedom. Namely, this is a consequence of such fully polarised initial state wavefunctions, which has nothing to do with final states, and hence no dependence on photon energies.

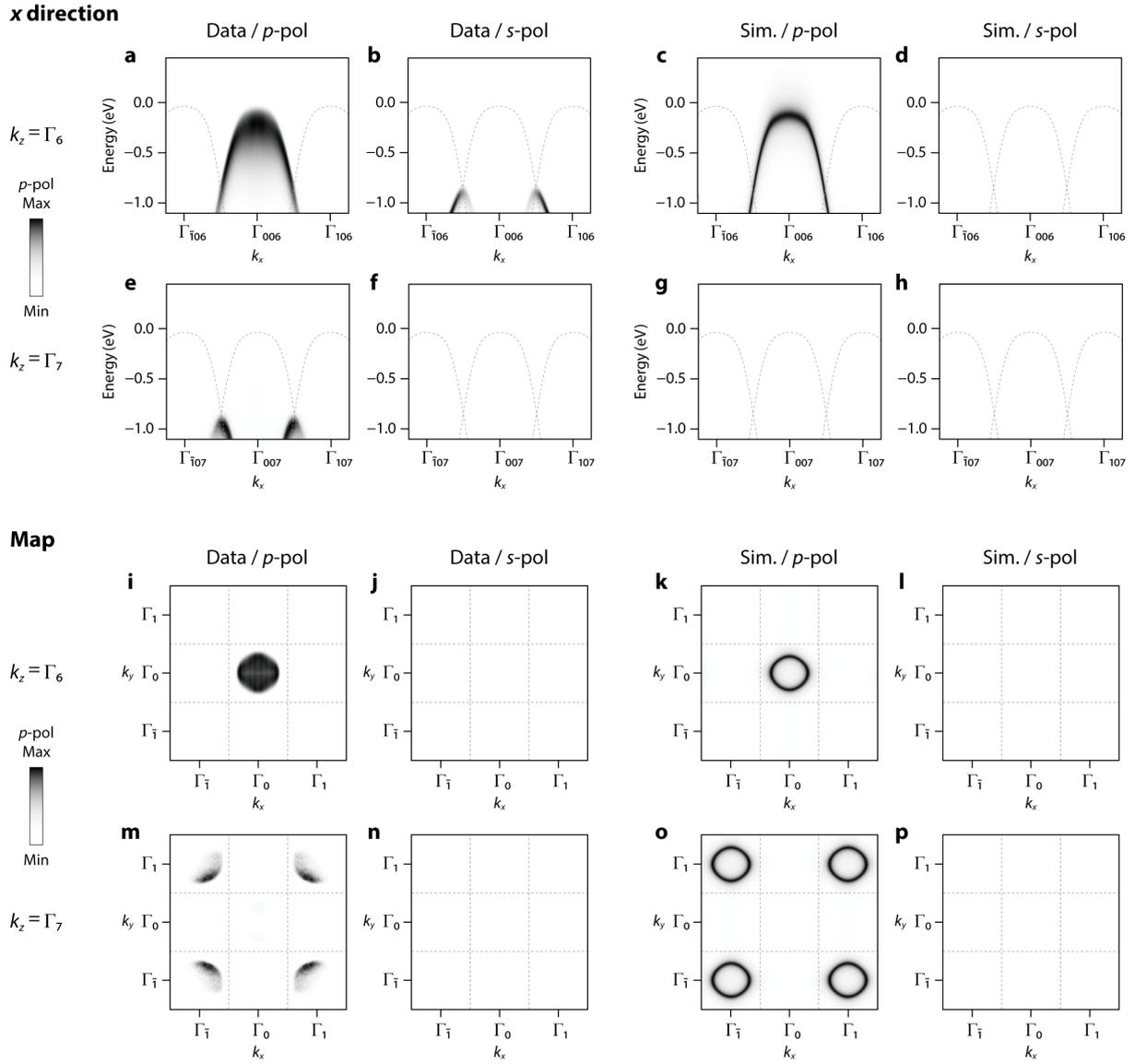

**Extended Data Fig. 6 | ARPES data of PdSe$_2$ compared with simulations. a-p,** ARPES data and simulations of band dispersion in $k_x$ (**a-h**) and constant energy maps at $E = -0.4$ eV (**i-p**). The ARPES data or simulations and light polarisation are indicated by Data / p-pol (**a,e,i,m**), Data / s-pol (**b,f,j,n**), Sim. / p-pol (**c,g,k,o**), and Sim. / s-pol (**d,h,l,p**), as marked on the top. The photon energy is 87 eV for $k_z = \Gamma_6$ (**a,b,i,j**) and 119 eV for $k_z = \Gamma_7$ (**e,f,m,n**), as marked on the left. After background normalisations, the same colour scale is applied to each pair of data taken by p-polarised light and s-polarised light. Grey dotted lines show the expected location of valence bands in **a-h** and zone boundaries in **i-p**. The bright and dark states can be reproduced by spectral simulations based on the model of sublattice interference.

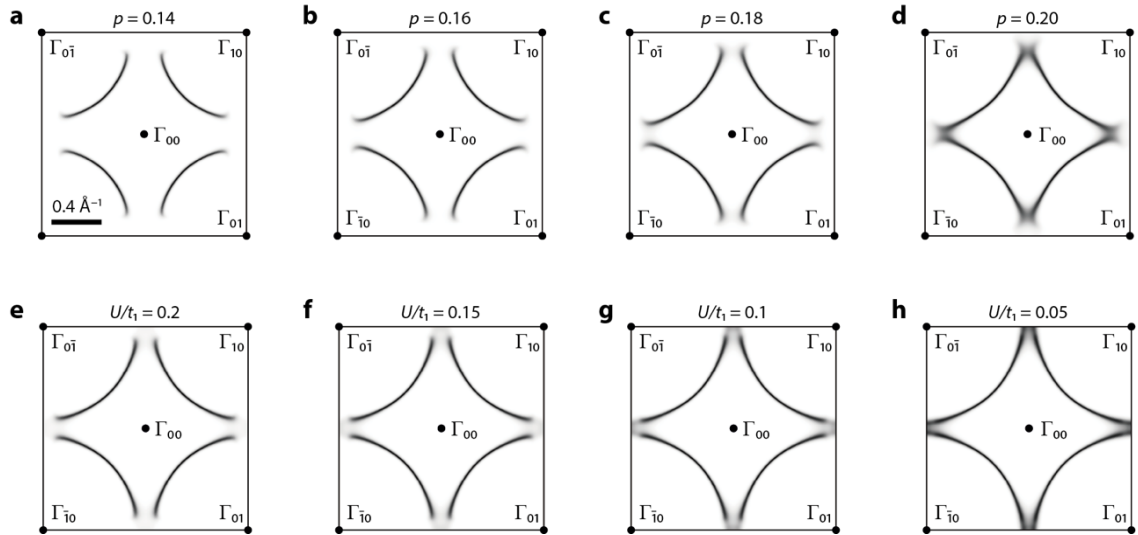

**Extended Data Fig. 7 | Doping and temperature dependence of Fermi arcs. a-d,** Fermi surfaces of single-layer cuprates simulated with our model (Methods) at $U = 0.4t_1$ by shifting $E_F$ for the hole concentration ($p$) marked on top of each panel. **e-h,** Fermi surfaces of single-layer cuprate simulated with our model at $p = 0.17$ by varying $U/t_1$ as marked on top of each panel. Our model of sublattice interference reproduces key aspects of phenomenology in doping and temperature dependence of Fermi arcs[31,33]: Those in **a-d** reproduce a key feature in doping dependence that the length of Fermi arcs grows in size with $p$. More importantly, the anti-nodal gap reduces in magnitude with increasing either doping or temperature. This can also be well reproduced by our simulations in **e-h** by reducing $U/t_1$ that represents the strength of symmetry breaking, no matter whether its origin is CDW or antiferromagnetism.

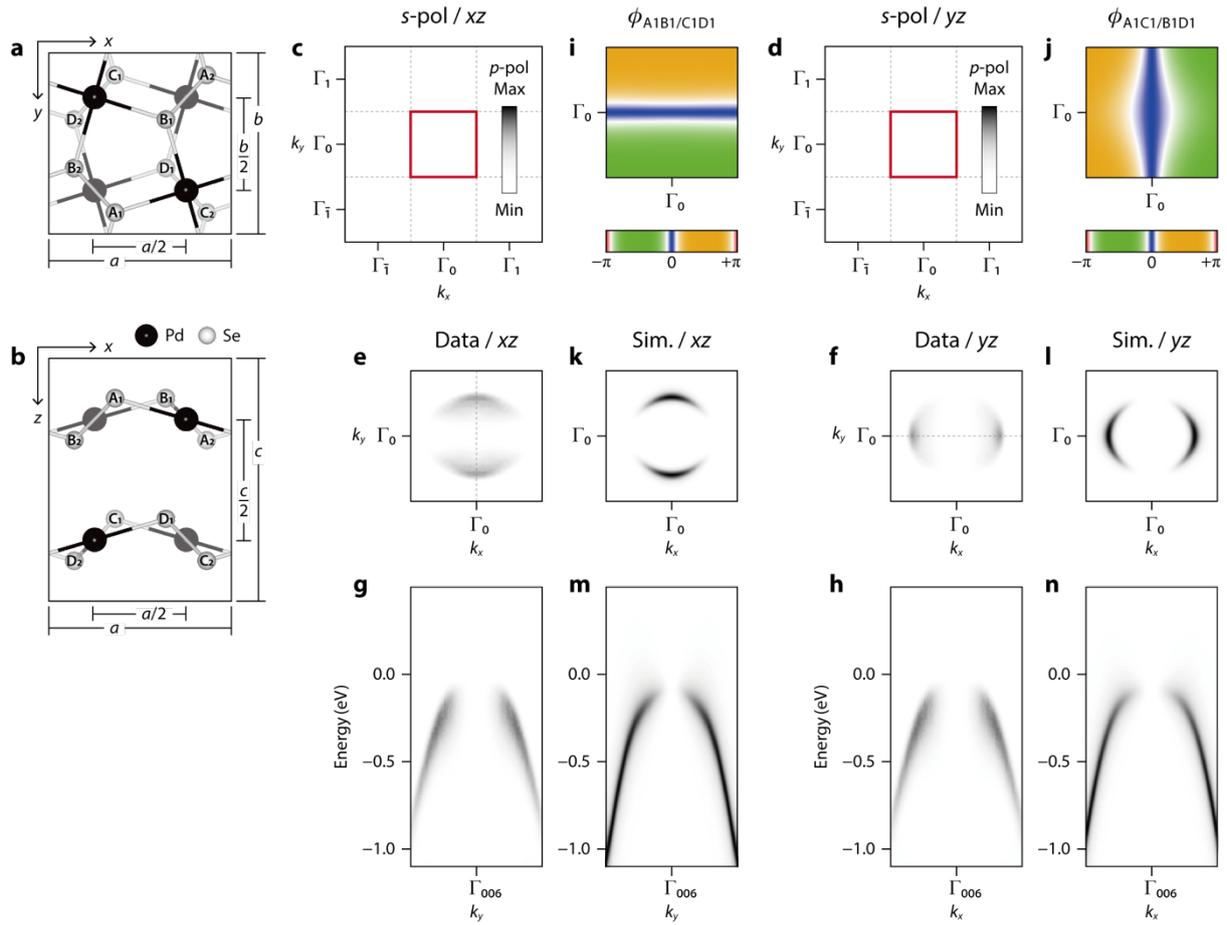

**Extended Data Fig. 8 | Se contributions in the ARPES data of PdSe$_2$. a,b,** Crystal structure of PdSe$_2$ illustrated by the ball-and-stick model and viewed from the top (**a**) and the side (**b**). The 8 Se atoms in the primitive cell are labelled A1–D1 and A2–D2. **c,d,** ARPES data of PdSe$_2$, taken by the *s*-polarised light of 87 eV ($k_z = \Gamma_6$) at the scattering plane of *xz* (**c**) and *yz* (**d**), and plotted at $E = -0.4$ eV as a function of $k_x$ and $k_y$. **e,f,** Part of constant-energy maps marked in **c,d** by red squares and shown in the extremely narrow colour scale. **g,h,** ARPES data of PdSe$_2$ shown in $k_y$ (**g**) and $k_x$ (**h**) along the grey dotted lines in **e,f**, respectively. **i,j,** Relative phases of $\phi_{A1B1/C1D1}$ (**i**) and $\phi_{A1C1/B1D2}$ (**j**) obtained from tight-binding models and shown at $k_z = \Gamma_{2n}$ for the same area as in **e,f** ($\phi_{A1B1/C1D1} = \phi_{A2B2/C2D2}$, $\phi_{A1C1/B1D2} = \phi_{A2C2/B2D2}$). The colour scale in units of radian is given at the bottom. **k-n,** Corresponding simulations of constant-energy maps (**k,l**) and band dispersions (**m,n**) with relative phases in **i,j** and sublattice interference. There is overall little ARPES intensity on the valence band of PdSe$_2$ taken by *s*-polarised light in **c,d**. However, if narrowing down the intensity scale for the area marked by the red square, we found there is a remnant feature with the intensity pattern that vanishes at $k_y = 0$ along the $k_x$ axis in **e** and at $k_x = 0$ along the $k_y$ axis in **f**. Even these characteristic intensity patterns can be reproduced well by ARPES simulations based on the model of quantum interference between 8 Se atoms (see Methods), as shown in **k-n**. This is essentially because the relative phase of parity pairs connected with respect to the scattering plane is 0 (blue) at $k_y = 0$ along the $k_x$ axis in **i** and at $k_x = 0$ along the $k_y$ axis in **j**, which cannot be excited by *s*-polarised light.

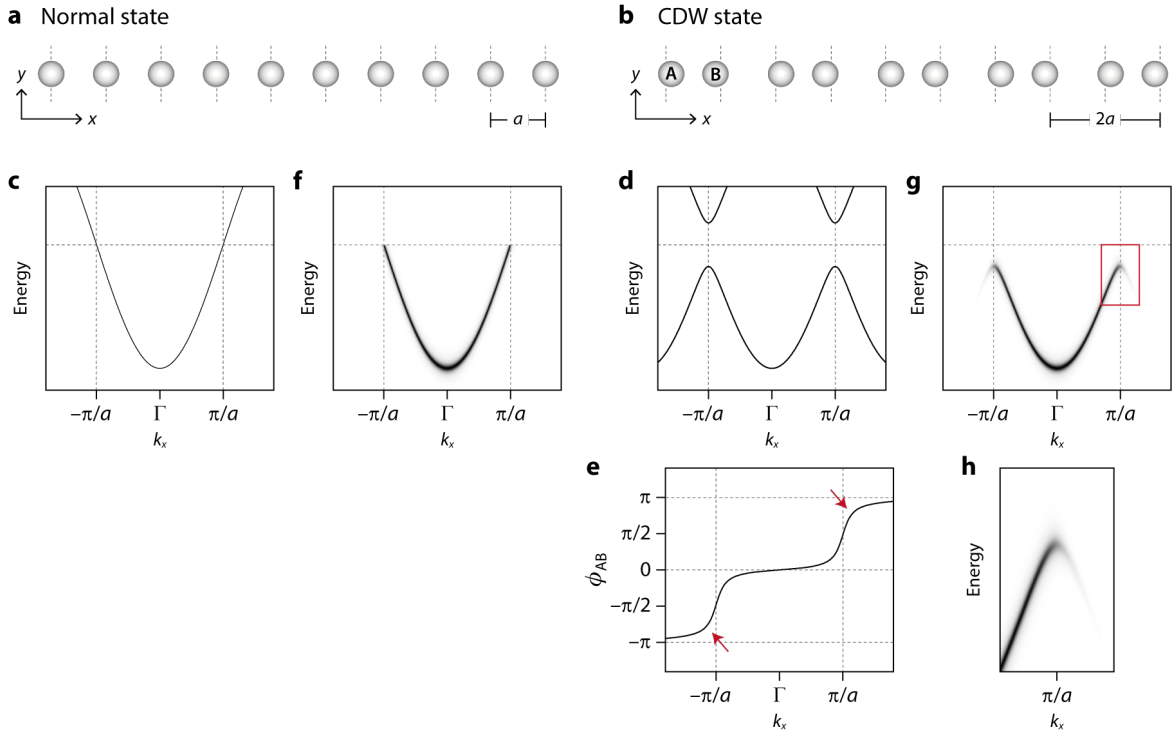

**Extended Data Fig. 9 | ARPES simulation for charge density waves. a,b,** Crystal structure of monoatomic chains running along *x* in normal (**a**) and CDW (**b**) states. The two sublattices formed by dimerization are labelled A and B in **b**. **c,d,** Band structure of monoatomic chains in normal (**c**) and CDW (**d**) states calculated by tight-binding models and plotted along $k_x$. **e,** $\phi_{AB}$ of the lower energy band in **d** obtained from tight-binding models and shown along $k_x$. **f,g,** ARPES simulations for normal (**f**) and CDW (**g**) states based on the model of sublattice interference. With the scattering plane of *xz*, $M^\mathbf{k}$ can be written in the form of $1 + e^{i\phi_{AB}}$ arising from integration over the scattering plane. This is independent of light polarisation because no parity conversion is expected across this scattering plane. The Fermi energy is assumed to be at the grey horizontal dotted lines. **h,** Magnified view of ARPES simulations marked by the red box in **g**, which shows typical spectral features that gradually diminish in the vicinity of energy gaps or along back folding of the main band. This can be explained by incomplete phase polarisations near $\pm\pi/a$ indicated by red arrows in **e** as a natural consequence of the two sublattices located slightly off the half-translation positions in **b**. Those centred at $\pm 2\pi/a$ with $\phi_{AB} \approx \pm\pi$ cannot be detected by ARPES, because they are nearly the dark states.